\documentclass[a4paper, aip, jap, reprint, superscriptaddress]{revtex4-1}

\usepackage{color}
\usepackage{graphicx}
\usepackage{amsmath}

\renewcommand{\Re}{\mathrm{Re}}
\newcommand{\WHz}{\mathrm{W/\sqrt{Hz}}}
\newcommand{\mum}{\mathrm{\mu m}}
\newcommand{\mus}{\mathrm{\mu s}}
\newcommand{\ee}[1]{\times 10^{#1}}

\newcommand{\qp}{\mathrm{qp}}
\newcommand{\opt}{\mathrm{opt}}
\newcommand{\pb}{\mathrm{pb}}
\newcommand{\bb}{\mathrm{bb}}
\newcommand{\dark}{\mathrm{dark}}
\newcommand{\rd}{\mathrm{read}}
\newcommand{\res}{\mathrm{res}}

\begin{document}

\title{Equivalence of Optical and Electrical Noise Equivalent Power of Hybrid NbTiN-Al Microwave Kinetic Inductance Detectors}

\author{R.M.J. Janssen}
\email{r.m.j.janssen@tudelft.nl}
\affiliation{Kavli Institute of Nanoscience, Faculty of Applied Sciences, Delft University of Technology, Lorentzweg 1, 2628CJ Delft, The Netherlands}

\author{A. Endo}
\affiliation{Kavli Institute of Nanoscience, Faculty of Applied Sciences, Delft University of Technology, Lorentzweg 1, 2628CJ Delft, The Netherlands}
\affiliation{Department of Microelectronics, Faculty of Electrical Engineering, Mathematics and Computer Science (EEMCS), Delft University of Technology, Mekelweg 4, 2628CD Delft, The Netherlands}

\author{P.J. de Visser}
\affiliation{SRON Netherlands Institute for Space Research, Sorbonnelaan 2, 3584CA Utrecht, The Netherlands}
\affiliation{Kavli Institute of Nanoscience, Faculty of Applied Sciences, Delft University of Technology, Lorentzweg 1, 2628CJ Delft, The Netherlands}

\author{T.M. Klapwijk}
\affiliation{Kavli Institute of Nanoscience, Faculty of Applied Sciences, Delft University of Technology, Lorentzweg 1, 2628CJ Delft, The Netherlands}
\affiliation{Physics Department, Moscow State Pedagogical University, Moscow, 119991, Russia}

\author{J.J.A. Baselmans}
\affiliation{SRON Netherlands Institute for Space Research, Sorbonnelaan 2, 3584CA Utrecht, The Netherlands}

\begin{abstract}
We have measured and compared the response of hybrid NbTiN-Al Microwave Kinetic Inductance Detectors (MKIDs) to changes in bath temperature and illumination by sub-mm radiation. We show that these two stimulants have an equivalent effect on the resonance feature of hybrid MKIDs. We determine an electrical NEP from the measured temperature responsivity, quasiparticle recombination time, superconducting transition temperature and noise spectrum, all of which can be measured in a dark environment. For the two hybrid NbTiN-Al MKIDs studied in detail the electrical NEP is within a factor of two of the optical NEP, which is measured directly using a blackbody source.
\end{abstract}

\maketitle
\begin{figure*}
	\centering
	\includegraphics[width=1.0\textwidth,keepaspectratio]{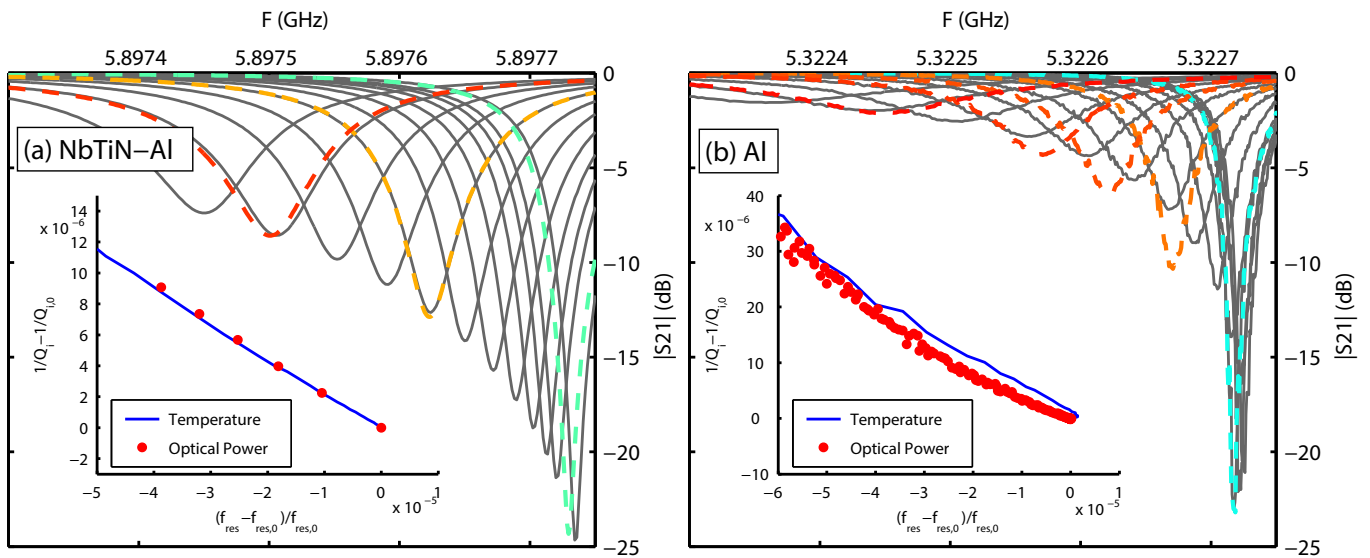}
\caption{(a) The evolution of the resonance curve as a function of increasing temperature (black) and increasing optical loading (color) of a hybrid NbTiN-Al MKID. We see that this evolution is identical, as can be seen from the relative change in resonator losses, $(1/Q_i-1/Q_{i,0})$, as a function of the change in resonance frequency, $(f_{\res}-f_{\res,0})/f_{\res,0}$ (inset). (b) As (a), but for a fully Al CPW MKID\cite{deVisser2014b}. Here a difference in the effect temperature and sub-millimeter radiation have on the complex conductivity is observed.}
\label{Figure:Gao3}
\end{figure*}
\begin{figure*}
	\centering
	\includegraphics[width=1.0\textwidth,keepaspectratio]{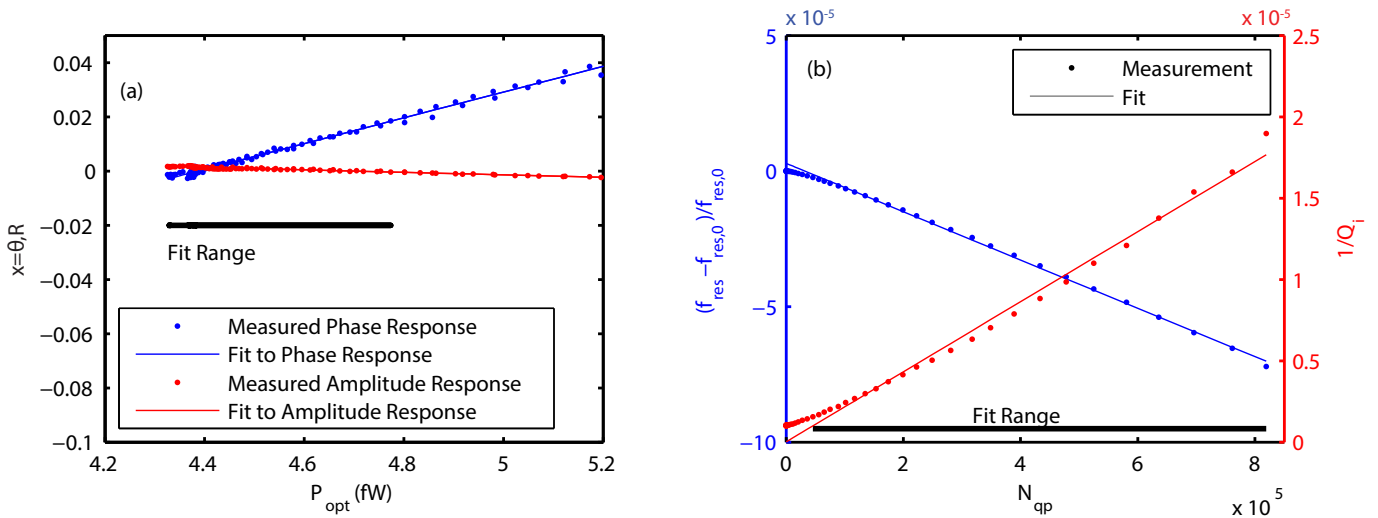}
\caption{(a) The measured change in phase (blue dots) and amplitude (red dots) as a function of the absorbed sub-millimeter radiation. From this measurement the phase (blue line) and amplitude (red line) responsivity is determined using a linear fit. (b) The measured (dots) change in resonance frequency (blue) and resonator loss (red) as a function of the number of thermally generated quasiparticles as given by Eq. \ref{quasiparticles}. A linear fit (lines) to this data gives us a quantity, which is proportional to the temperature responsivity (see Eq. \ref{y2x}).}
\label{Figure:Fits}
\end{figure*}

In the development of megapixel sub-millimeter cameras for ground-based astronomy two different implementations of the Microwave Kinetic Inductance Detector (MKID)\cite{Day2003} are currently being pursued. One implementation is the Lumped Element MKID (LEKID)\cite{Doyle2008} made from TiN\cite{LeDuc2010}. The high normal state resistance of TiN allows direct photon absorption and a lower read-out frequency without a dramatic increase in pixel size\cite{McKenney2012}. A lower readout frequency reduces the cost of read-out electronics. However, LEKIDs made from this high resistivity material have shown an anomalous optical response\cite{Gao2012,Bueno2014}. An alternative MKID implementation is the lens-antenna coupled hybrid NbTiN-Al MKID\cite{Yates2011}, which integrates an Al absorber in a NbTiN resonator. Sub-mm radiation creates quasiparticles in the Al, which will be trapped there, because the superconducting gap of NbTiN is much larger than that of Al. The lens-antenna coupled hybrid MKIDs have shown the expected photon noise limited performance in both phase and amplitude readout down to 100 fW of optical loading as well as a high optical efficiency\cite{Yates2011,Janssen2013}.\\
A Noise Equivalent Power (NEP) in the $10^{-19}$ $\WHz$ range has been measured electrically for MKIDs\cite{LeDuc2010,deVisser2012}. The electrical NEP  is determined from the MKIDs temperature responsivity, quasiparticle recombination time, superconducting energy gap and noise spectrum, all of which can be measured in a dark environment. Based on a simplified model-analysis Gao et al.\cite{Gao2008c} have argued that the change in complex conductivity due to thermally and optically excited quasiparticles is equivalent. This would imply that the electrical NEP is a convenient alternative to a full optical evaluation, which requires a time-consuming measurement and a dedicated setup with a controlled illumination source\cite{Janssen2013}. However, the relationship presented by Gao et al.\cite{Gao2008c} is not universally applicable to MKIDs, as illustrated in Fig. \ref{Figure:Gao3}(b). Fig. \ref{Figure:Gao3}(b) compares the response to thermal and optical excitations of well-studied Al coplanar waveguide (CPW) MKIDs\cite{deVisser2014b}. Clearly, the temperature response deviates significantly from the optical response.\\
In this Letter we report a thorough analysis of the optical and electrical NEP of lens-antenna coupled hybrid NbTiN-Al MKIDs. In Fig. \ref{Figure:Gao3}(a) we show that, unlike the fully Al CPW MKIDs, hybrid MKIDs have an identical response to optical illumination and a change in bath temperature. We define a conversion between temperature and optical power based on energy arguments and show that for these hybrids the electrical NEP, which is determined from the temperature response, quasiparticle recombination time, superconducting energy gap and noise spectrum, is within a factor of two of the directly measured optical NEP.\\
The hybrid NbTiN-Al MKIDs\cite{Janssen2013} we study are quarter wavelength CPW resonators (length $\sim 5$ mm), which consist of two sections. In the first section, at the open end of the resonator, the CPW is wide and made from NbTiN. In the second section, which is approximately 1 mm in length and located at the shorted end of the resonator, the NbTiN CPW is narrow and the central line is made from Al instead of NbTiN. This hybrid MKID design simultaneously maximizes the responsivity and minimizes the two-level system (TLS) noise\cite{Gao2007} contribution.
\\
We measure the properties of two representative devices (numbered No. 1 and No. 2) using a pulse tube pre-cooled adiabatic demagnetization refrigerator with a box-in-a-box cold stage design\cite{Baselmans2012b}. In this design the 4-by-4 array of MKIDs is fully enclosed in a 100 mK environment with the exception of a 2 mm aperture, which is located 15 mm above the approximate center of the MKID array. The aperture is isotropically illuminated by a large blackbody\cite{deVisser2014b}. Metal mesh filters define a 50 GHz bandpass around a central frequency of 350 GHz. The passband is matched to the antenna design \cite{Janssen2013}. This allows us to create an unpolarized illumination over a wide range of powers. From the blackbody temperature, $T_{\bb}$, the filters and the optical coupling between the blackbody and the MKIDs we can determine the absorbed photon power, $P_{\opt}$, to within 6\%\cite{Janssen2013}. The magnetic field strength of the ADR is used to control the temperature, $T$, of the sample. Magnetic shielding prevents these fields from entering our sample stage. In our experiment we use a bath temperature $T_0 = 100$ mK and a blackbody temperature $T_{\bb,0} = 4.2$ K  as our initial condition for both our optical and thermal measurement. Quantities measured at these initial conditions will be denoted with subscript zero. Starting from $(T_{\bb,0},T_{0})$ we change either $T_{\bb}$ or $T$ in the optical and thermal measurements, respectively. All measurements are made at a fixed readout power $P_{\rd} = -80$ dBm. In general the effect of microwave power on a quasiparticle distribution, which is modified from our baseline by an elevated temperature or optical illumination, is different.\cite{Goldie2013,Guruswamy2014} However, we expect that for the same number of quasiparticles, generated either thermally or optically, the effect of readout power is comparable.\\
In order to determine the optical NEP of our detectors we first determine the base temperature resonance frequency of the MKID, $f_{\res,0}$, and we measure at $f_{\res,0}$ the noise spectrum in phase readout, $S_{\theta,0}$, and amplitude readout, $S_{A,0}$. Second, we measure the optical responsivity of the device, $\delta x / \delta P_{\opt}$, by monitoring $x=\theta,A$ at $f_{\res,0}$, while increasing $P_{\opt}$. Fig. \ref{Figure:Fits}(a) shows the measured phase (blue dots) and amplitude (red dots) response as a function of $P_{\opt}$. We determine the optical responsivity directly from this measurement by a linear fit to the measured response at $P_{\bb} \leq 1.1P_{\bb,0}$. These fits are shown in Fig. \ref{Figure:Fits}(a) for phase readout (blue line) and amplitude readout (red line) and their slopes give the optical responsivity values presented in Table \ref{Table:Respons}. Table \ref{Table:Respons} also lists the uncertainty of the measured optical responsivity. The uncertainty is 5-10\% and the result of uncertainty in the fit ($\sim 4$\%) and the uncertainty in $P_{\opt}$ ($\sim 6\%$).\\
An electrical NEP can only be used as a proxy for the optical NEP, if the response to temperature and sub-mm radiation is equivalent. Fig. \ref{Figure:Gao3}(a) shows the evolution of the resonance curve of a hybrid NbTiN-Al MKID as a function of increasing temperature (solid black lines) and increasing optical loading (dashed colored lines). For every optical loading a temperature can be found, which shows an identical resonance feature. This shows that, in case of the hybrid MKIDs, temperature generates an identical response in the MKIDs as sub-mm radiation. This shows sub-mm radiation and thermal excitation modifies the quasiparticle energy distribution in the Al in a comparable way and corroborates our expectation that alterations due to $P_{\rd}$ are a second order effect, which is below the measurement accuracy. This is further illustrated by the inset of Fig. \ref{Figure:Gao3}(a), which shows the change in internal quality factor, $Q_i$, as a function of the change in resonance frequency, $f_{\res}$, for changing $T$ (solid blue line) and $P_{\opt}$ (red dots). In contrast to hybrid MKIDs, radiation induces less losses for the same frequency shift in fully Al CPW MKID\cite{deVisser2014b} as shown in Fig. \ref{Figure:Gao3}(b).
\begin{table}[b]
\begin{tabular}{|l||c|c|}
\hline
MKID No. 1 & Optical & Electrical \\ \hline
$d\theta/dP$ & $(47.5\pm2.0)\ee{12}$ & $(35.5\pm6.3)\ee{12}$ \\ 
$dA/dP$ & $(-4.59\pm0.29)\ee{12}$ & $(-4.29\pm0.70)\ee{12}$ \\ \hline
MKID No. 2 & Optical & Electrical \\ \hline
$d\theta/dP$ & $(32.7\pm3.5)\ee{12}$ & $(20.3\pm3.6)\ee{12}$ \\ 
$dA/dP$ & $(-3.39\pm0.37)\ee{12}$ & $(-2.41\pm0.43)\ee{12}$ \\ \hline
\end{tabular}
\caption{The optical and electrical phase and amplitude responsivity as well as their uncertainty for the two MKIDs.}
\label{Table:Respons}
\end{table} \\
We determine the electrical (dark) responsivity\cite{Baselmans2008}, $\delta x/\delta P_{\dark}$, which based on energy arguments we expect to be equivalent to the optical responsivity, by
\begin{equation}
\frac{\delta x}{\delta P_{\dark}} = \frac{\eta_{\pb}\tau_{\qp}(T)}{\Delta(0)} \frac{\delta x}{\delta N_{\qp}(T)}
\label{eqrespons2}
\end{equation}
Here $\delta x/\delta N_{\qp}(T)$ is the temperature responsivity, $\tau_{\qp}$ is the quasiparticle recombination time, $\Delta(0)$ is the BCS superconducting energy gap and $\eta_{\pb}$ is the pair breaking efficiency. Table \ref{Table:Respons} gives the measured electrical responsivity and their uncertainty for our MKIDs. In the paragraphs below we detail how each of the parameters required for the calculation of $\delta x/\delta P_{\dark}$  as well as their individual measurement uncertainty is obtained.\\
At temperatures $T\ll T_c$ Mattis-Bardeen theory\cite{MB1958} predicts a linear relation between the number of thermal quasiparticles and the real and complex part of the conductivity\cite{Gao2008c} or equivalently the internal losses and MKID resonance frequency. Fig \ref{Figure:Fits}(b) shows the measured (dots) linear relation between the number of quasiparticles, $N_{\qp}(T)$, and the resonance frequency (blue) or the resonator losses (red). We determine $N_{\qp}(T)$ from the bath temperature, $T$, using
\begin{equation}
N_{qp}(T) = V \times 4N_0 \int_0^{\infty} \ N_s(E,T) f_{FD}(E,T) dE
\label{quasiparticles}
\end{equation}
Here $V \approx 135$ $\mum^3$ is the volume of the Al in the MKID, $N_0=1.7\ee{10}$ $\mum^{-3} \ \mathrm{eV^{-1}}$ the single spin density of states of Al at the Fermi energy\cite{Parks1969}, $f_{FD}(E,T)$ is the fermi-dirac energy distribution and $N_s(E,T)$ the normalised BCS quasiparticle density
\begin{equation}
N_s(E) = \Re\left(\frac{E}{\sqrt{E^2-\Delta^2(T)}}\right)
\label{QPdos}
\end{equation}
From a linear fit to temperatures $0.18T_c < T < 0.25T_c$ we obtain the temperature responsivity, $\delta y/\delta N_{qp}$, of the resonance frequency, $y = (f_{\res}(T)-f_{\res,0})/f_{\res,0}$, and internal quality factor, $y=1/Q_i$. The fit range and resulting function are shown in Fig \ref{Figure:Fits}(b). From $\delta y/\delta N_{qp}$ we obtain the temperature responsivity, $\delta x/\delta N_{qp}$, for phase, $x=\theta$, and amplitude, $x=A$.
\begin{subequations}
\label{y2x}
\begin{align}
        \frac{\delta \theta}{\delta N_{\qp}} &= \frac{-4Q}{f_{\res,0}}\frac{\delta (f_{\res}-f_{\res,0})}{\delta N_{\qp}},\label{freq2phase}\\
        \frac{\delta A}{\delta N_{qp}} &= -2Q\frac{\delta (1/Q_i)}{\delta N_{\qp}}, \label{Qi2Amp}
\end{align}
\end{subequations}
Here $Q$ is the measured (total) resonator quality factor.\\
By selecting $T > 0.18 T_c$ as a fit range for the temperature response we avoid the region below $N_{qp} \approx 0.4\ee{5}$ where the non-linear TLS response dominates over the quasiparticle response, as shown by the measurements in Fig \ref{Figure:Fits}(b). At $T > 0.18 T_c$ our minimum blackbody temperature, $T_{\bb,0}=4.2$ K, which corresponds to $P_{\opt,0} = 4.5$ fW, generates a negligible amount of quasiparticles compared to those generated thermally. The commonly used approximation for Eq. \ref{quasiparticles} (for example Eq. 7 by Gao et al.\cite{Gao2008c}) systematically underestimates $N_{\qp}(T)$ by up to 5\% at $T=0.25T_c$. As a result we would overestimate $\delta x/\delta N_{\qp}$ by 5.6\%. To eliminate this error we use the full BCS integral.\\
A quasiparticle recombination time $\tau_{\qp,0} = 138 \pm 20$ $\mus$ is determined\cite{Janssen2013} from the roll-off in the noise spectrum. This method is ideal, because one observes the MKID in an equilibrium situation. However, this can only be done if photon noise\cite{Yates2011} or generation-recombination noise is observed\cite{deVisser2011}. Alternatively, the recombination time can be obtained by measuring the MKID's response to short high energy pulses.\cite{deVisser2011,Note1}\\ 
We determine the superconducting energy gap using $\Delta(0)=1.76k_bT_c$. We find a $T_c=1.283\pm0.019$ K for the Al used in the hybrid MKIDs. This $T_c$ is the mid-point value from four-point DC measurements of the film resistance, $R(T)$, as a function of temperature. The DC measurement uses Al Hall bars of 3 and 100 $\mum$ wide, fabricated simultaneously with the MKIDs. We observe a change in $T_c$ between the two Hall bars. Inspection of a 4" wafer excludes any effects due to changes in Al thickness, but shows a variation of 0.5 um in the width of lines with the same design, due to the wet etch of the Al. We infer a 1.5\% uncertainty in $T_c$ due to these lithographic variations. We do not understand the physical mechanism responsible for the observed $T_c$ variation. Nevertheless, the uncertainty in $T_c$ is important, because it introduces an exponential uncertainty in the temperature responsivity through Eq. \ref{quasiparticles}. A change of 1.5\% in $T_c$ introduces a 10\% change in $\delta x/\delta N_{\qp}$.\\
The pairbreaking efficiency is usually taken to be\cite{Kozorezov2000,Kurakado1982} $\eta_{pb} = 0.57$. Recent simulations by Guruswamy et al.\cite{Guruswamy2014} of the absorption of sub-millimeter radiation by superconductors has corroborated this value for thick films of a BCS superconductor. However, they have also shown that $\eta_{pb}$ depends on the device materials, geometry and the photon energy. For the hybrid NbTiN-Al MKIDs in this experiment, which uses 48 nm thick Al\cite{Janssen2013} and receives photons with an energy $h\nu \approx 7.5\Delta(0)$, the simulations show an $\eta_{\pb}\approx 0.4$. Accordingly, we have used $\eta_{\pb}= 0.4$ in our analysis.
\begin{figure}
\centering
\includegraphics[width=1.0\columnwidth]{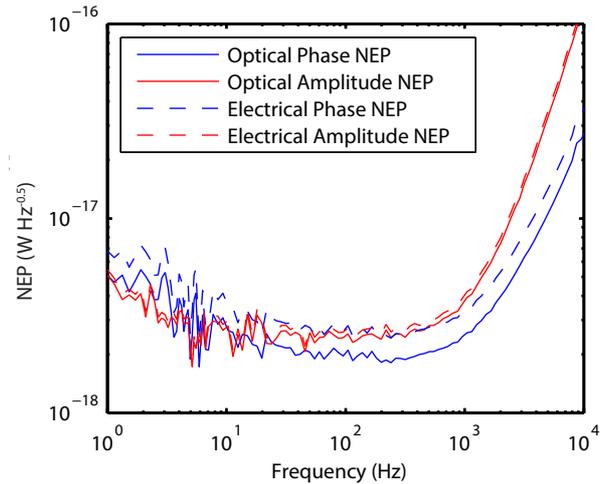}
\caption{The optical NEP (solid line) compared to the electrical NEP (dashed line) for phase (blue) and amplitude (red) readout. The spectral shape of the electrical and optical NEP is identical, because they are both based on the noise spectrum measured at the base temperatures $T_{\bb,0} = 4$ K and $T_{0} = 100$ mK.}
\label{Figure:Match}
\end{figure}\\
Fig. \ref{Figure:Match} shows the optical (solid) and electrical (dashed) NEP for phase (blue) and amplitude (red)  readout for KID No. 1. The NEPs are determined from the measured noise spectra and responsivities. The spectral shape of the optical and electrical NEP is identical, because they are both based on the noise spectrum measured at the base temperatures. Therefore, the difference between the optical and electrical NEP level is entirely due to the difference in the optical and electrical responsivity. The measured responsivities are given in Table \ref{Table:Respons}. It is clear from Fig. \ref{Figure:Match} that the electrical NEP coincides with the optical NEP to within 37\% for phase readout and within 7\% for amplitude readout. For MKID No. 2 the optical and electrical NEP are within 62\% and 41\%, respectively.\\
With the exception of the amplitude responsivity of MKID No. 1, the difference between the measured optical and electrical NEP is larger than the measurement uncertainties. $\Delta(0)$ and $\eta_{\pb}$ are the most likely candidates causing this discrepancy. We anticipate that the responsivity discrepancy is the result of not knowing the exact value of $\Delta(0)$, which we obtain from the mean-field $T_c$. However, the BCS factor of 1.76 used in this conversion is a theoretical quantity obtained in the limit of weak electron-phonon coupling. It is known to increase for a more realistic electron-phonon coupling, although for aluminium we do not expect a large deviation. A factor of 1.85 would bring all responsivities to within $1.1\sigma$. Only, a direct measurement of $\Delta(0)$ would resolve this uncertainty. $\eta_{\pb}$ can resolve the discrepancy to a similar degree. This would require a phonon lifetime to phonon trapping time ratio four times larger than literature values.\cite{Guruswamy2014,Kaplan1976,Kaplan1979}\\
The sensitivity to the exact value of $\Delta_0$ and $\eta_{\pb}$ as well as the already large $(\sim 20$\%) uncertainties in the electrical NEP makes clear that it is impossible to determine the optical efficiency by the comparison between the optical and electrical NEP. A very reliable way to determine the optical efficiency is to use the photon noise limited NEP as described in detail by Janssen et al.\cite{Janssen2013}.\\
A possible equivalence between the optical and electrical NEP, and therefore responsivity, is based on the assumption that illumination by sub-millimeter radiation and providing an elevated temperature creates the same change in the weighted spatial average of the complex conductivity measured by the resonator. As shown in Fig. \ref{Figure:Gao3} for a hybrid MKID the evolution of the resonance feature is identical for temperature and optical loading. However, in a full Al MKID radiation induces less losses for the same frequency shift\cite{deVisser2014b}. We interpret this difference as a result of the hybrid MKID geometry in which only the short aluminum section acts as the absorber. The 1 mm Al section is long enough to absorb the incoming radiation, but is shorter than the quasiparticle diffusion length. In addition, the electric field is roughly constant over this section of the MKID, which means that over the whole length of the aluminum the quasiparticle-density has an identical contribution to the responsivity, regardless of its position.\\
Consequently, we show that for the specific case of hybrid MKIDs it is justified to assume the electrical NEP to be identical to the optical NEP. The geometrical advantage of hybrid NbTiN-Al CPW MKIDs is also present in fully Al LEKIDs\cite{Mauskopf2014}, where both the radiation absorption and current are uniform within the inductive end of the resonator. This advantage is absent in Al CPW MKIDs. For fully TiN MKIDs, as mentioned in the introduction, it appears to be impossible to rely on an electrical NEP, as defined in this Letter, as a measure for the optical performance, because the response to radiation is in many aspects anomalous and possibly related to the inhomogeneous nature of the superconducting state\cite{Bueno2014}.\\
In conclusion, we have shown for hybrid NbTiN-Al MKIDs that 
\begin{enumerate}
\item the weighted spatial average of the complex conductivity measured by hybrid MKIDs is the same for thermal and optical excitation.
\item the electrical NEP, which is determined from the temperature responsivity, quasiparticle recombination time, superconducting transition temperature and noise spectrum, is within a factor of two of the optical NEP, which is measured directly using sub-millimeter radiation.
\end{enumerate}
We argue that this is the result of the specific implementation of the hybrid NbTiN-Al MKID, and that in different MKID embodiments the equivalence between optical and electical response is not a priori justified.
\begin{acknowledgments}T.M. Klapwijk and R.M.J. Janssen are grateful for support from NOVA, the Netherlands Research School for Astronomy, to enable this project. A. Endo is grateful for the financial support by NWO (Veni grant 639.041.023) and the JSPS Fellowship for Research Abroad. T.M. Klapwijk acknowledges financial support from the Ministry of Science and Education of Russia under contract No. 14.B25.31.0007 and from the European Research Council Advanced grant No. 339306 (METIQUM).
\end{acknowledgments}

%
\end{document}